\documentclass[12pt]{article} \usepackage{latexsym}
\usepackage{amsfonts} \usepackage{amsmath,cite, graphics}
\newcommand{\sm}{{\mathsf s}}
\begin{document}

\title{Visualizing the Mass and Width Spectrum
of Unstable Particles}

\author{N.L.~Harshman\\
Formerly: Department of Physics and Astronomy\\
Rice University\\ Houston, TX\\
Currently: Department of Computer Science, Audio Technology and
Physics\\  American University\\ Washington, DC}

\maketitle

\begin{abstract}

\vspace{1cm}

\normalsize

Several graphical representations of the mass and width spectrum of
unstable subatomic particles are presented.  Such plots are useful
tools for introducing students to the particle zoo and provide
students an alternate way to organize conceptually what can seem like an
overwhelming amount of data.  In particular, such graphs highlight
phenomenological features of unstable particles characteristic of 
different energy and time scales.

\end{abstract}

\vspace{1cm}

\section{Introduction}

When confronting the vast array of subatomic particles detected and created in
particle physics experiments, new students (and old researchers) may
become overwhelmed with the volume of descriptive information.  The purpose
of this note is to present a visual tool to supplement other
schemes used to organize the diversity of unstable particles.
The idea is simple: what does a plot of width versus mass of subatomic
particles look like and what information does it convey about the
kinematics and dynamics of these particles?
I hope that instructors of particle physics may find it useful when
introducing new visitors to the particle zoo and that practitioners
may find an alternate perspective stimulating.

The most useful feature of these graphs
is not that they summarize information about the spectrum or
structure of unstable particles, but that they highlight the
distinctions between the phenomenological 
signatures of unstable particles at different energy and time scales.
These graphs provide a natural way of grouping together
particles, not by their physical content, but by how they are
produced, how they decay, and how their characteristic parameters are
measured.  This alternate taxonomy distinguishes three classes of
unstable particles: decaying
states, which decay via the weak interaction and whose exponential
decay rate can be measured, resonance states, which decay primarily
through the strong interaction and are measured as features in the
cross section, and a middle class of ``platypus'' states, whose
instability parameters require more subtle measurements to determine.
Also, these 
graphs can be used to call attention to how the limitations and
prejudices of past experiments affect what is known about the mass and
width spectrum of particles.

\section{Which Unstable Particles?}

Figure 1 depicts the mass $M$ and width $\Gamma$ of 139 unstable
particles, with mass plotted logarithmically on the horizontal
axis and width plotted logarithmically vertically.  The shape of
each plotted point indicates the type of particle it is (gauge boson,
e.g.)
and for the hadrons the style (black, gray, or hollow) 
indicates some information about the quark content.

The data for the 139 unstable particles come from the 2002 edition of
\emph{The Review of 
Particle Physics}~\cite{PDG}, and in particular the list of well-known,
reasonably well-measured
unstable particles from the ``Summary Tables of Particle Properties''
therein.  Not every particle in the ``Summary Table'' has been
included; only those particles found in the file \cite{PDGMC} that the Particle
Data Group tabulates, of the mass and width data,
for use in
Monte Carlo event generators and detector simulators.  For unstable
particles whose lifetimes $\tau$ are quoted in the
\emph{Review}~\cite{PDG} and not their widths,
the width values in the Monte Carlo file are found using
the Weisskopf-Wigner relation $\Gamma = \hbar/\tau$ (more will be said
about this later).

The list of particles from the Monte Carlo file \cite{PDGMC} has been
modified and applied in the following way in Fig.~1 and subsequent figures:
\begin{enumerate}
\item The stable particles, the proton, electron,
photon and neutrinos, are excluded.
\item The nearly-stable neutron is neglected for reasons of scale.
\item The Monte Carlo
file includes some particles for which only an upper bound of the
width has been measured.  They 
have been excluded.  Examples include some light
unflavored meson resonances like the $f_0(980)$, other meson
resonances such as the $D_s^{*\pm}$ and $\chi_{b0}(1P)$, and a few
baryon resonances like the $\Sigma_c(2520)^+$.
\item The top quark, not truly an independent particle like the others
in the list, is not included.
\item The symbol plotted for a particle also represents its
antiparticle, except for the neutral K-mesons.  For these, the mass
eigenstates $K^0_S$ and $K^0_L$ are plotted instead of
the flavor eigenstates $K^0$ and $\bar{K}^0$.

\item A single symbol is plotted for all different charge-species of a
baryon unless different masses for different charges have
been measured.  For example, each point representing a $\Delta$ baryon
represents all four charge species $\{++, +, 0, -\}$ corresponding to
quark contents $\{uuu, uud, udd, ddd\}$.

\end{enumerate}

Then what unstable particles \emph{are} included?  The weak gauge bosons W
and Z are at the high energy extreme and 
the muon is at the low energy extreme.  The other unstable lepton, the
tau, is in 
the middle, 
along with a host of hadrons made up of five out of the six quarks:
up, down, 
strange, charm and bottom.  While the gauge bosons and leptons are
to our best knowledge structureless; the hadrons are composite.
Subsequent references to particles refer just to this set of
well-established, well-measured unstable particles, and therefore should
not be taken to refer to all possible particles that have or have not
been observed or theorized.

Looking at Figure 1, it may be tempting to ask if there is a
functional dependence of the width on the mass.  In principle the
widths of unstable particles are calculable from the masses of the
quarks and leptons and other Standard Model parameters, although in
practice such calculations are difficult or impossible, especially for
hadrons.  For the purposes of this article, the width and mass are
considered independent phenomenological parameters to be determined
from experiment.

Figure 1 does make apparent the general trend
of increasing 
width with increasing mass, which is explained by phase space effects.
The decay 
rate is roughly proportional to the phase space of the decay products
and the more massive the unstable particle, the more decay channels
are available.

To better elucidate the properties of the scattered distribution in
Figure 1, several 
other partitions or sections of the data are included below.  Figure 2
plots the masses of the 139 unstable particles in order of increasing
mass, i.e.~ in rank order from the least massive to the most massive.
Figure 3 is similar, except it plots them in rank order of increasing
width.  Figure 
4 plots the unitless ratio of width to mass in increasing order.
These different graphs give further clues about the structure of the
mass and width spectrum and how to identify phenomenologically-similar
groups of unstable particles.

\section{Mass Spectrum}

A striking feature of Figure 1
is that while the widths run over a range of 18 orders of magnitude,
the masses are constrained within three decades, with most
of them between 1 GeV and 10 GeV.  This fact says more about the types
of experiments that have been performed than about the ``essential
nature'' of the mass 
spectrum.  Far more particle searches have focussed on this energy range
for a variety of historical and practical reasons.

Another perspective on the mass spectrum can be gained from Figure 2.
Qualitatively, gaps in the plot indicate how quark
composition affects (and effects) the mass spectrum
of hadrons.  
For example, at the low end, twelve of the first thirteen points are mesons
consisting of the 
three light quarks: up, down and strange.  The non-meson among the
thirteen is the muon, the lightest unstable particle, which
historically was mistaken for a meson. 
The twelve mesons have 
masses less than the 
lightest (undepicted) baryons, the proton and neutron at about 940 MeV.
There is a gap between the first three points, which represent the muon,
$\pi^0$ and $\pi^\pm$, and the next set, which include the lightest
mesons composed of strange quarks, the $K^\pm$, $K^0_S$, $K^0_L$ and
$\eta$.

At the high end of the mass spectrum, except for the gauge bosons, the
points are dominated by hadrons containing heavy quarks.  The lightest
hadron containing a charm quark, the $D^0$, is eighty-fifth on the list
with a mass of 1865 MeV and the lightest hadrons containing a bottom
quark are the B-mesons $B^\pm$ and $B^0$ with ranks of 127 and 128 and
masses of 5279 MeV.  Similar to the strange
quark mass threshold jump seen at the lower energies, here there are slight
jumps at 117, the lightest particle with two charm
quarks $\eta_c$, at 127, the lightest particle with one bottom quark
$B^\pm$, and at 132, the lightest particle with two bottom quarks 
$\Upsilon(1S)$.

The fact that above number 109 all points
represent hadrons containing charm or bottom quarks does not mean that
unstable particles containing only light quarks are not found in this
mass range. 
It only indicates that such resonances have not been the focus of
experimental searches in that energy range.  Also, at such
high energies production of light mesons is so copious that wide, light-meson
resonances get lost in the background.  

Finally, the true heavy-weights, the gauge bosons, with masses near 100
GeV, stand alone at the far sides of Figure 1 and Figure 2.  Their
isolation is again an artifact of experimental particle physics
history, not of some fundamental nature of the mass spectrum of
particles.  To
make precise measurements of these particles entailed building
experiments with far more energy and at a far greater cost.  To better
explore the physics at this scale and to search for potentially heavier
particles such as the Higgs boson will require the next generation of
particle accelerators.

\section{Width Spectrum}

Unlike the mass spectrum, the width spectrum (i.e., the plot of the
 widths of the 139 unstable particles in increasing order) depicted in
 Figure 3 is
not as dependent on the energy scales of experiments that have been performed.
Figure 3 spans many decades of the width, and even the most massive
particles can have very small or very large widths.  For example, the
$\Lambda_b$ baryon, with mass and width $(5.62\ \mbox{GeV}, 5.36\times
10^{-13}\ \mbox{GeV})$, has a mass rank of 130 out of 139, but a width
rank of 15.  It is
considered unlikely that we will discover any more stable or extremely
long-lived subatomic particles, so future discoveries will likely either fit
into this graph or, perhaps for new ultra-massive particles like the
Higgs boson, be appended to
the end, beyond the gauge bosons.  Therefore the shape of this graph is
unlikely to change much as unstable particles are added, whereas new
particle discoveries will probably 
smooth out the high energy, quark-mass threshold gaps in the mass
spectrum Figure 2. 

Looking at Figure 3, it seems natural to roughly break the graph into
three parts: particles with widths $\Gamma < 10^{-8}$ MeV, $10^{-5}$ MeV $<
\Gamma < 10$ MeV, and $\Gamma > 10$ MeV.  This division becomes even
more sensible when Figure 4, which lists the particles in order
of the width-to-mass ratio $\Gamma/M$, is considered.  The first 40
particles in both 
Figures 3 and 
4 have nearly the same order; only after that is there a substantial
reshuffling.  In Figure 4 a heuristic division between nearly the same three
classes of particles could be made at $\Gamma/M < 10^{-10}$, $10^{-8}
< \Gamma/M < 10^{-2.5}$ and $\Gamma/M > 10^{-2.5}$.

\subsection{First Class: Decaying States}

The interesting thing about these classes is that they have physical
significance.  In Figure 1, this first class of particles is the arc
of leptons, 
mesons and baryons 
along the lower part of the graph. From the point of view of
fundamental interactions, what these particles
have in common is that they decay via the weak interaction, and
consequently they are long-lived.

Long-lived is of course a relative
term, but widths $\Gamma < 10^{-8}$ MeV correspond to lifetimes $\tau >
10^{-14}$ s.  In terms of phenomenology, that means for particles in this
first class the exponential decay rate can be measured directly.  This
measurement involves finding the distance traveled 
between the production location and the decay vertex.  This information,
combined with kinematic information of mass, momentum, energy and/or
speed, can be used to find the time-of-flight in the rest frame of the
particle.  A histogram of all the time-of-flights for a certain
particle type can be fit to an exponential to get the lifetime.

Long life makes this first class seem like the most 
``particle-like'' of all the unstable particles. In many calculations
they can be
approximated as stable because they are stable with
respect to the much shorter time scales dictated by the strong and
electromagnetic interactions.  Their width/mass ratios $\Gamma/M$ are
so small that any mass uncertainty can be neglected in kinematic
calculations.

\subsection{Third Class: Resonances}

In contrast, particles in the third class have widths
$\Gamma > 10$ MeV, corresponding to particles with lifetimes
$\tau < 10^{-22}$, and to a width-to-mass ratio $\Gamma/M$ greater than a
part in a thousand.  Because of this mass uncertainty, they can be
produced in experiments with energies 
substantially lower or higher than their central (quoted) mass value.
This class of particles decays primarily through strong 
interactions and the vast lifetime difference separating
them from the 
first class is attributable to the strength of the strong interaction
and the massiveness of the
weak gauge bosons.

It
may seem unjustified to consider these short-lived unstable particles as the
same kind of object as long-lived unstable particles, so vast are the
differences in their instability parameters.  The
distinction between these two classes is sometimes codified in the
language of particle physics: the long-lived particles are
called decaying states and the short-lived are
called resonances.  This 
distinction arises because of the different ways that decaying states
and resonances are observed experimentally.

Particle resonances are detected as rapid variations (usually peaks) in the
cross section.  As the 
center-of-mass energy of a collision is scanned over some range, there
may appear an enhancement of the elastic cross section 
or the cross section into a particular set of inelastic channels.
After extracting the background and accounting for uncertainties in
the preparation and detection apparatuses and other effects (such as
radiative corrections), the resonant cross section $\sigma_R$ as a function of
center-of-mass energy (or center-of-mass energy squared $\sm$) can be
extracted.  This process can become more complicated if there are
multiple resonances in the same energy region, interfering resonances,
or background-resonance interference.

Neglecting these complications,
the resonance cross
section $\sigma_R$ (or resonance lineshape or linewidth) can be fit to a
theoretical function 
and the mass and width 
extracted.  Typically, the function used is the Breit-Wigner or
Lorentzian function, which can be parameterized in terms of mass and
width as
\begin{equation}
\sigma(\sm) \propto \left|\frac{1}{\sm - (M - i\Gamma/2)^2}\right|^2,
\end{equation}
or by several other parameterizations.   This functional form
for the resonance lineshape can be derived by 
associating the resonance to a pole in 
the scattering S-matrix~\cite{GW,Bohm}.  Alternate functions exist;
for example, in 
perturbative quantum field theory, the on-mass shell renormalization
scheme leads to a different definition for mass and
width (for a 
discussion of this as applied to the Z-boson, see \cite{NPB}).  
For some resonances, such as the $\Delta$ baryons, the
\emph{Review}~\cite{PDG} cites values for mass and width corresponding
to both the Breit-Wigner and perturbation theory
definitions.

To measure the resonance cross section accurately and to extract a value
for mass and 
width, whichever function and parameterization are used, requires
experimental energy 
resolution precise enough to trace out the lineshape.  Practically,
this means the energy resolution should be smaller than the ratio
$\Gamma/M$ for that particular resonance.  As a result, the lineshapes
of particles in
the first class, with $\Gamma/M < 10^{-10}$, cannot be measured in
this way.
Measurements of width and lifetime are 
physically very distinct processes that apply to phenomena at very
different energy 
scales.

The width and lifetime of a
particle are related theoretically by the 
Weisskopf-Wigner formula $\tau = \hbar/\Gamma$, which was originally
proposed in the context of atomic electronic lineshapes~\cite{WW}.  This
relation is considered so standard as to appear to be a definition or
identity or direct consequence of the uncertainty principle, but to the best
knowledge of this author, this relationship has never been verified
experimentally in
the regime of subatomic physics~\cite{wwexp}.  The Wigner-Weisskopf relation
between the width and decay rate or inverse lifetime $\Gamma/\hbar$ is
derived as an approximation in non-relativistic scattering in many
textbooks (for example \cite{GW, WWder}), and can be proven identically in the
non-relativistic and relativistic case for the mathematical
object called a Gamow vector~\cite{Gamow1}.

\subsection{Middle Class: Troublemakers}

The middle class of particles on the chart, with widths roughly between
$10^{-5}$ MeV $<
\Gamma < 10$ MeV and width-to-mass ratios $10^{-8}
< \Gamma/M < 10^{-2.5}$, are awkwardly placed from an experimental point
of view.  Their lifetimes are so short that direct measurement of
exponential decay is extremely difficult or impossible; their widths
are so narrow 
that few experiments have the energy resolution to accurately trace their
lineshape.  While the
middle class is easily distinguishable from the decaying states in
Figure 1, they blend into the bottom of the resonance states.  There
is no clear gap characteristic of the difference in strength between 
electromagnetically- and strongly-decaying states.

Physically, what do states in this middle class have in common?
Some of these
states have substantial branching ratios into 
electromagnetic decay channels like the $\pi^0$ and $\Sigma^0$ and do
not decay via the
strong interaction.  Some, like the $D^*(2020)^\pm$ are barely above the 
energy threshold of for their primary decay channels and their decays
are therefore 
phase-space suppressed.  All the mesons in
this class (except the $D^*(2020)^\pm$) are unflavored, i.e.\ composed
of quarks and antiquarks of the same flavor (or superpositions of
same-flavor quark/antiquark pairs) and are not energetic enough to
decay into strongly-favored channels.

Neither decaying states nor resonances by the definitions described
above, how are the widths of these states measured?  Below, the
experimental determinations of some  of these particles' widths are
described to give a flavor for some of the other tools at the disposal
of particle
physicists.

\begin{itemize}

\item With a lifetime of $(8.4 \pm 0.6) \times 10^{-17}$ s, the
exponential decay rate of the $\pi^0$ is very hard to measure.  The
$\pi^0$ decays almost exclusively into a pair of photons via the
electromagnetic interaction.  A
precision experiment by Atherton, et al.~\cite{Atherton}, fit an
exponential to three 
data points extracted from the distance between the production and the
decay of the
$\pi^0$ into 
two photons.  The measurement agrees (within error) with another
method used to measure widths and lifetimes: the Primakoff effect~\cite{Prima}.
Bombarding heavy nuclei with gamma rays, an interaction between the
incoming photon and a virtual photon can produce the $\pi^0$; this
process is called photo-production.  One can measure the cross section
for this process and relate it theoretically via the transition
amplitude to the partial width or decay rate
$\Gamma_{2\gamma}$  for
the process
$\pi^0 \rightarrow \gamma\gamma$.  From this, the total width is
determined by dividing the partial width $\Gamma_{2\gamma}$ by the
independently-measurable branching ratio $B_{2\gamma}$  into 
the two photon decay channel, i.e., $\Gamma =
\Gamma_{2\gamma}/B_{2\gamma}$.

\item The Primakoff effect has also been
used to measure the lifetime of the $\Sigma^0$ baryon
as $(7.4 \pm 0.7)\times 10^{-20}$ s.  The $\Sigma^0$ decays almost
exclusively to $\Lambda \gamma$.

\item For the $c\bar{c}$ resonances $J/\psi$ and  $\psi(2S)$ and the
$b\bar{b}$ resonances $\Upsilon(1S)$, $\Upsilon(2S)$ and
$\Upsilon(3S)$ the
partial width $\Gamma_{ee}$ is extracted from the \emph{integrated} cross
section, for example, in the elastic process $e^+e^- \rightarrow J/\psi
\rightarrow e^+e^-$.  Then the ratio of the elastic cross
section to the total cross section independently provides the  branching
ratio $B_{ee}$.  The total width (as above) is $\Gamma = \Gamma_{ee}/B_{ee}$.

\item Both the above method and the Primakoff effect have been applied
to the $\eta$ meson giving consistent results and a value for the
width of $\Gamma = 1.18 \pm 0.11$ keV. 

\item Finally, the $\mathrm{D}_0^*(2010)^\pm$, with a width of $\Gamma
= 96 \pm 4 \pm 22$ keV, has a mass just above the threshold for its
main decay channels $D^0\pi^\pm$ and $D^\pm\pi^0$.  Its width cannot be
measured directly, but can be extracted from fitting to simulations
of the the energy
distribution of decay products~\cite{D0(2010)}.

\end{itemize}

\section{Summary and Further Considerations}

The mass-width spectrum in Figure 1 does not
reveal as much information about unstable particles 
as Hertzsprung-Russell diagrams reveal about stellar composition and
evolution~\cite{astro}.  These graphs do not contribute to finding a
perturbative, 
renormalizable, elegant theory for predicting masses and widths of
hadrons based on standard model parameters.
Nonetheless, Figure 1 and subsequent partitions of it do show some structures
that correspond to significant phenomenological features.  As a result,
I think these graphs provide an excellent tool (or at least a starting
point) for instructing
students of particle physics about a host of physical phenomena and
experimental procedures.  A partial list of these features or ideas
would include:

\begin{itemize}
\item Relative strengths of the fundamental interactions.
\item Connection of the transition amplitude and scattering matrix to
measurable quantities.
\item Experimental measurement of exponential decay.
\item Resonance production and decay.
\item Cross section measurements.
\item Phase space dependence of width and decay rate.
\end{itemize}

Additionally, not just the viewing of these graphs, but the production of
these kinds of plots would be an excellent exercise for students,
undergraduate or 
graduate.  While familiarizing themselves with the particle zoo, they
could also practice using graphing and data
management software.  As an example of other kinds of plots possible,
Fig.~5 depicts the width versus mass of just the mesons on a linear
scale.

Finally, visualizing the mass and width spectrum of unstable
particles with such graphs makes the task of conceptually organizing the
physical data of hundreds of unstable particles a little easier.  It
provides an alternate way to group them into
phenomenologically-meaningful families, complimenting standard organization
schemes according to flavor and/or quark content.  Many physicists
have some biological, taxonomical part of their brains
to which I hope these graphs appeal.

\pagebreak

\pagebreak

\includegraphics{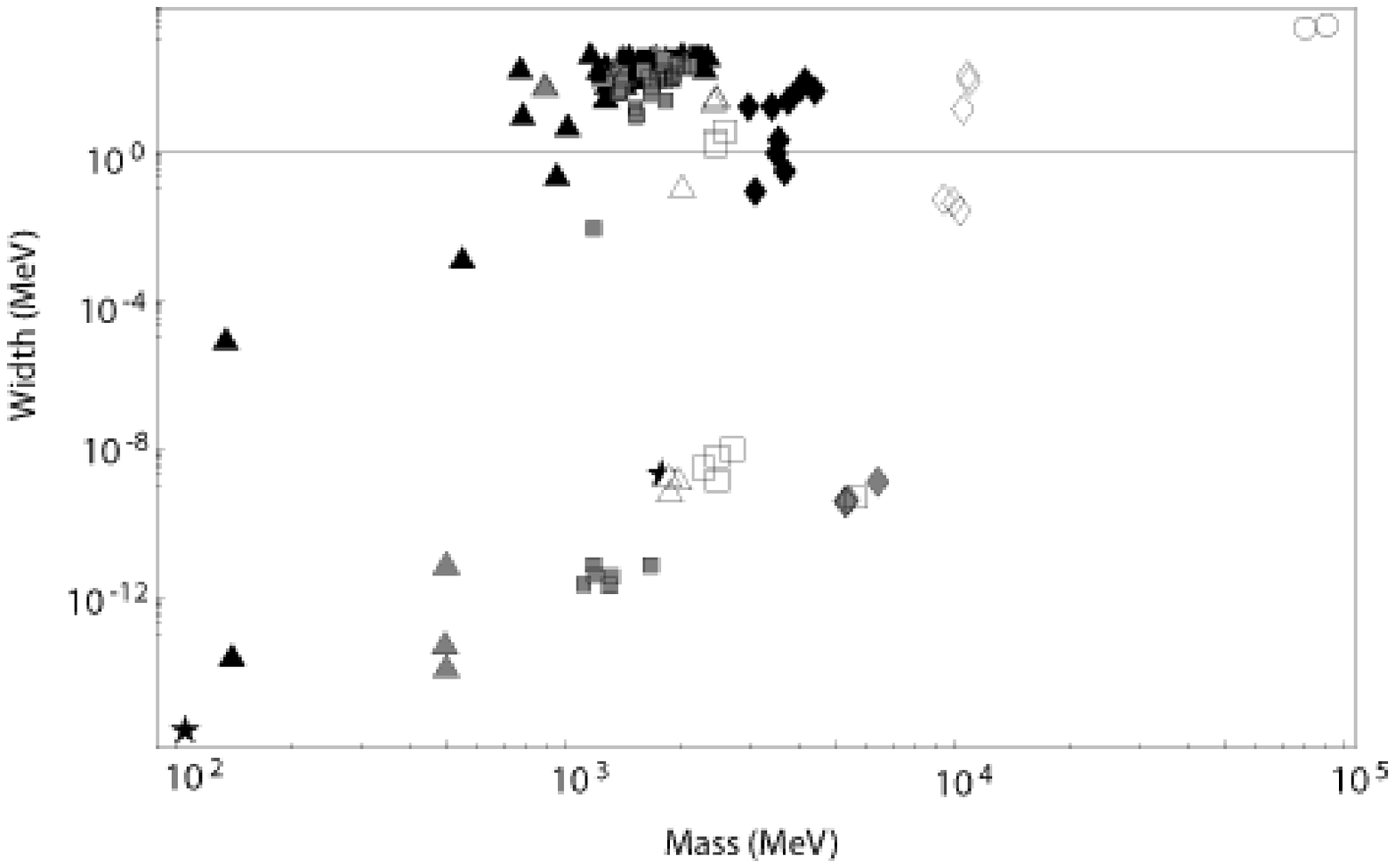}

\noindent Fig.~1. Log-log plot of Mass/MeV versus Width/MeV.  Choice
of 139 unstable
particles plotted described in text.  Key:
hollow circles---gauge bosons; black stars---leptons; black
triangles---light unflavored mesons; gray triangles---strange
mesons; hollow triangles---flavored charmed mesons (including
charmed-strange mesons); black diamonds---unflavored charmed
($c\bar{c}$) mesons; gray
diamonds---flavored bottom mesons (including 
bottom-strange and bottom-charmed
mesons); hollow diamonds---unflavored bottom ($b\bar{b}$) mesons; black
squares---$N$ and $\Delta$ baryons; gray 
squares---strange baryons (including $\Lambda$, $\Sigma$, $\Xi$ and
$\Omega$ baryons); hollow squares---charmed and bottom baryons.

\pagebreak
\includegraphics{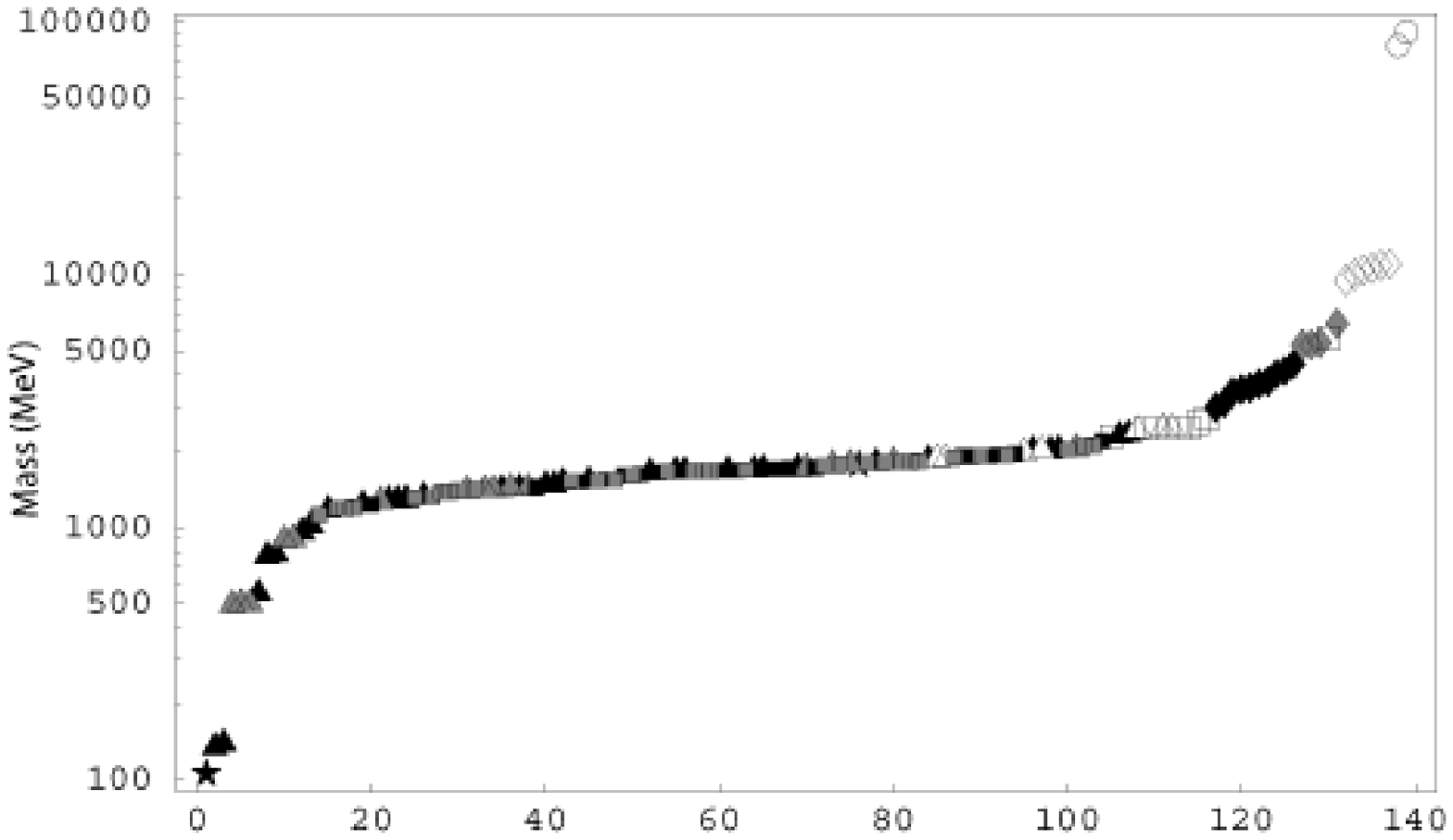}

\noindent Fig.~2.  Log-linear plot of Mass/MeV versus particle rank in
order of increasing
mass (out of 139 selected particles).  Key is same as in Fig.~1.

\pagebreak
\includegraphics{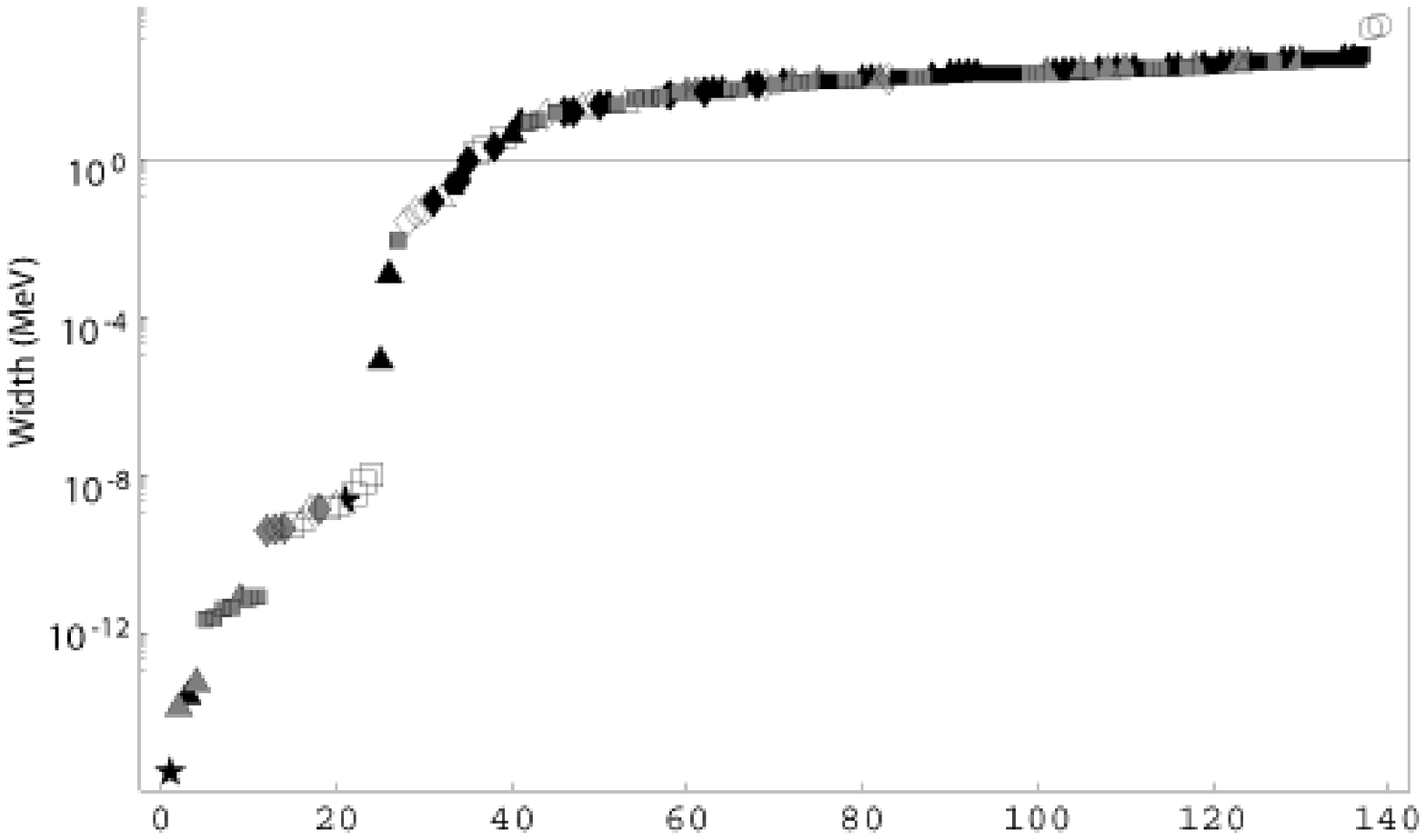}

\noindent Fig.~3.  Log-linear plot of Width/MeV versus particle rank in
order of increasing
width (out of 139 selected particles).  Key is same as in Fig.~1.

\pagebreak
\includegraphics{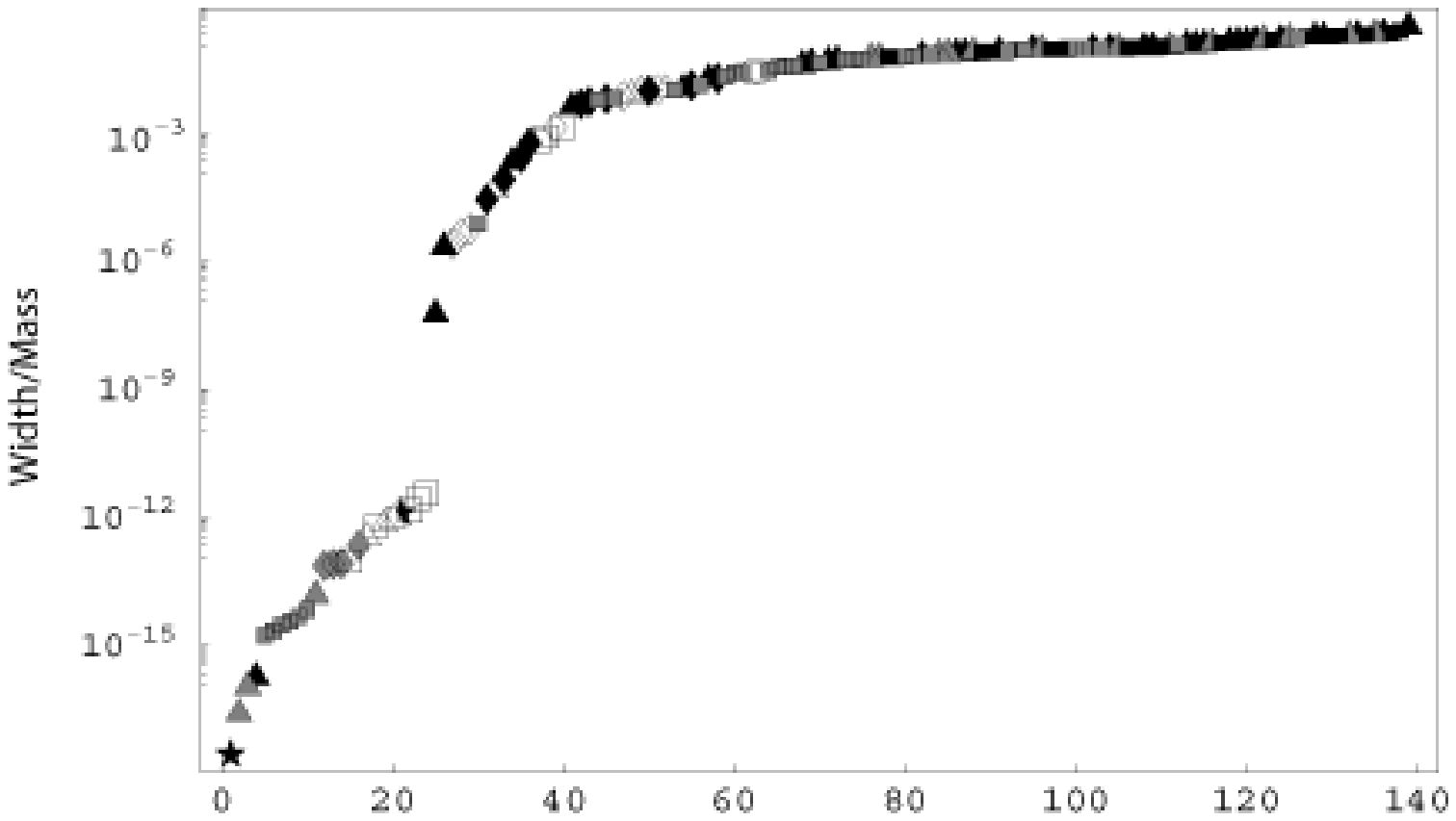}

\noindent Fig.~4.  Log-linear plot of width-to-mass ratio $\Gamma/M$
versus rank in
order of increasing width-to-mass ratio (out of 139 selected
particles).  Key is same as in Fig.~1. 

\pagebreak
\includegraphics{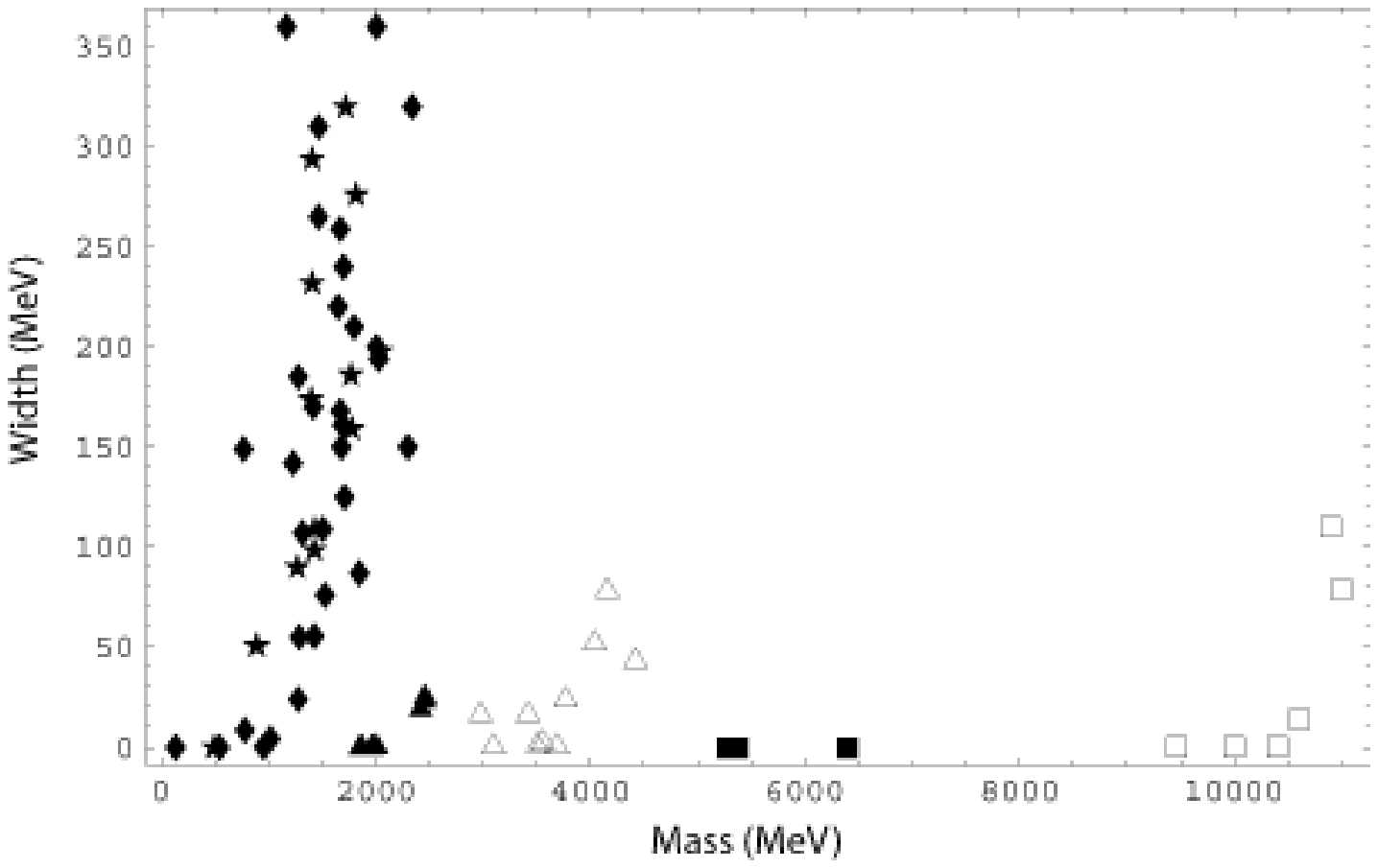}

\noindent Fig.~5.  Linear plot of width versus mass of unstable
mesons.  Key: black diamonds---light, unflavored mesons, black stars---strange
mesons; black triangles---charm-flavored mesons (including
charmed/strange-flavored mesons); hollow triangles---$c\bar{c}$-mesons;
black squares---bottom-flavored mesons (including
bottom/strange- and bottom/charmed-flavored
mesons); hollow squares---$b\bar{b}$-mesons.


\begin{thebibliography}{99}
\bibitem{wwexp} The Wigner-Weisskopf
relation has been verified some time ago for some nuclear unstable 
states~\cite{frau,ajp} using the  
M\"ossbauer effect to achieve the required energy resolution and more
recently for
one atomic unstable state\cite{Na} using 
ultra-cold atoms for the same purpose.
\end{thebibliography}
\end{document}